\documentclass[aps,twocolumn,showpacs,amsmath,amssymb,pra,reprint]{revtex4-1}

\usepackage{graphicx}%
\usepackage{dcolumn}
\usepackage{bm}
\usepackage{braket}
\begin{document}

\title{Analytical theory for the time-resolved  dynamical Franz-Keldysh effect under  circularly polarized light}
\author{T. Otobe}
\affiliation{$^1$National Institutes for Quantum and Radiological Science and Technology, Kyoto 619-0215, Japan}
\begin{abstract}
We report here the analytical formula for the time-resolved dynamical Franz-Keldysh effect (Tr-DFKE) under circularly polarized light. 
We assume the Houston function as the time-dependent wave function of the parabolic two-band system. 
Our formula shows that the sub-cycle change of the optical properties disappears, 
which is a significant feature of the Tr-DFKE under linear polarized light and is different from the static Franz-Keldysh effect.
\end{abstract}
\maketitle
\section{introduction}
In the last two decades, advances in laser sciences and technologies have led to intense coherent light with different characteristics becoming available.
 Ultra-short laser pulses can be as short as a few tens of an attosecond, forming the new field of attosecond science
\cite{atto01}.  
Intense laser pulses of mid-infrared (MIR) or THz frequencies  
have also become available recently \cite{HDBT11, Chin01}.  
By employing these extreme sources of coherent light, 
it is possible to investigate the optical response of materials in real time with a resolution much lower than an optical cycle 
\cite{atto01,Hirori11,Krausz13,Schultze13,Schultze14,Novelli13}.  
 
The dielectric function $\varepsilon(\omega)$ is the most fundamental  
quantity characterizing the optical properties of matter.  
The dielectric function observed in an ultra-fast  pump-probe experiment
should be  further considered as a  probe time ($T$) dependent function, $\varepsilon(T,\omega)$.
The modulation of the dielectric function  $\varepsilon(\omega)$  in the presence of  
electromagnetic fields has been a subject of investigation  
for many years. The change under a static electric field is known  
as the Franz-Keldysh effect (FKE)   
\cite{Franz58,Keldysh58,Tharmalingam63,Seraphin65,Nahory68,
Shen95,Sipe10,Sipe15}, 
and that under an alternating electric field is known as  
the dynamical FKE (DFKE)  
\cite{Yacoby68, Jauho96, Nordstorm98, Ajit04,Mizumoto06, Shambhu11}. 

Recently, we determined the sub-cycle change of the optical properties, i.e., time-resolved DFKE (Tr-DFKE),
which corresponds to the response and quantum path interference of a different phase locked dressed state \cite{otobe16}.
In particular, this ultra-fast change exhibits an interesting phase shift that depends on the field intensity and probe frequency. 
By utilizing this phenomenon, we can produce an ultra-fast modulator of light or an ultra-fast optical switch.

In previous work, we showed that the field intensity of the pump light and the band width of the probe pulse are crucial parameters \cite{otobe16}.
Another possible control parameter is the polarization of the light.
In this work, we present the analytical formula for the DFKE under a circularly polarized pump light.
We found that the time-dependent change in the optical properties completely disappears under a circularly polarized pump light.
Our formula and numerical results indicate that the response of the each dressed states is still important.

 The present organized  as follows.
 In Sec. II, we  will develop an analytical formulation for the Tr- DFKE under a circularly polarized light 
 by employing a parabolic two-band model.
 In Sec. III, we will present the numerical results and compare with them with the DFKE under a linearly polarized light and static FKE.
 In Sec. IV, a summary will be given.
\section{Formulation}
To derive the time-dependent conductivity, we will revisit a simple model that we reported in a previous work \cite{otobe16}.
The probe electric field is assumed to be weak enough to be treated  
using the linear response theory. We denote the electric current caused by  
the probe field as $J_p(t)$, which is assumed to be parallel to the
direction of the probe electric field. Thees are related by the time-domain  
conductivity $\sigma(t,t')$ as 
\begin{equation} 
J_p(t) = \int_{-\infty}^t dt' \sigma(t,t') E_p(t'), 
\label{def_sigma} 
\end{equation} 
where $E_p(t')$ is the electric field of the probe pulse.
We note that the conductivity $\sigma(t,t')$ depends on both 
times $t$ and $t'$ rather than the just time difference
$t-t'$ due to the presence of the pump pulse.

In the following developments, we will consider a simplified description:
electron dynamics in the presence of the pump and probe fields is 
assumed to be described by a time-dependent Schr\"odinger 
equation for a single electron,  
\begin{equation}
 i\frac{\partial}{\partial t} \psi_n(\vec r,t) 
= \left[\frac{1}{2m_e}\left(\vec p+\frac{e}{c}\vec A(t)\right)^2+V(\vec r) \right] \psi_n(\vec r,t) 
\label{TDSE}
\end{equation}
where $\psi_n(\vec r,t) $ is the time-dependent wave function of $n$-th band, 
 $\vec A(t)$ is the vector potential of the pump light field, and 
 $V(\vec r)$ is a time-independent, lattice periodic potential.
 In this paper, we employ the atomic units for all equations.
We express the solution of this equation using the time-dependent 
Bloch function $v_{n\vec k}(\vec r,t)$ as
$\psi_n(\vec r,t)=\sum_{\vec{k}}e^{i\vec k \vec r} v_{n \vec k}(\vec r,t)$, 
where $\vec{k}$ is the  Bloch wavevector.  

We further assume that in the presence of the pump field
described by a vector potential $\vec A_P(t)$,  the solution of 
Eq. (\ref{TDSE}) is well approximated by the  Houston 
function \cite{Yacoby68,Houston}. 
Using static Bloch orbitals  $u_{n\vec k}(\vec r)$, and orbital energies 
$\epsilon_{n\vec k}$  which satisfy
\begin{equation}
\left[\frac{1}{2m_e}\left(\vec p+\vec k \right)^2+V(\vec r) \right] u_{n\vec k}(\vec r)
= \epsilon_{n\vec k} u_{n\vec k}(\vec r),
\end{equation}
the Houston function can be expressed as
\begin{equation} 
 w_{n\vec{k}}(\vec{r},t)=u_{n \, \vec k_P(t)}(\vec{r}) 
 \exp \left[-i \int^t \epsilon_{n \,\vec k_P(t')} dt' \right],
 \end{equation} 
where $\vec k_P(t)$ is defined by $\vec k_P(t) = \vec k + e\vec A_P(t)/c$.

We will consider a circularly polarized electric field with the vector potential,
\begin{equation}
\vec{A}_P(t)=A_0(\cos \Omega t,\sin\Omega t,0).
\end{equation}
Since the pump electric field which is periodic in time, we have 
$\vec A_P(t+T_{\Omega}) = \vec A_P(t)$, where $T_{\Omega}$ is the period of
the pump field and is related to the frequency $\Omega$ by $T_{\Omega}=2\pi/\Omega$. 
The conductivity $\sigma(t,t')$ also has  the periodicity, 
\begin{equation}
\sigma(t,t')=\sigma(t-T_{\Omega}, t'-T_{\Omega}). 
\end{equation}
We will produce a Fourier expansion of $\sigma(t,t-s)$ which is
periodic in $t$ with the period $T_{\Omega}$,
\begin{equation}
\sigma(t,t-s) = \sum_{n=-\infty}^{\infty} e^{in\Omega t} \sigma^{(n)}(s),
\label{sigma_Fex}
\end{equation}
where $\sigma^{(n)}(s)$ is defined by
\begin{equation}
\sigma^{(n)}(s) = \frac{1}{T_{\Omega}} \int_0^{T_{\Omega}} dt
e^{-in\Omega t} \sigma(t,t-s).
\label{def_sigma_n}
\end{equation}
The time-resolved frequency-dependent conductivity $\tilde\sigma^I(T_p,\omega)$
for an impulsive probe field can be expressed as
\begin{equation}
\tilde{\sigma}^I(T_p,\omega) =
\sum_n e^{in\Omega T_p} \tilde{\sigma}^{(n)}(\omega+n\Omega),
\label{td_sigma}
\end{equation}
where $\tilde\sigma^{(n)}(\omega)$ is the Fourier transform of
$\sigma^{(n)}(s)$.
For a general probe field of
\begin{equation}
E^p(t)=f_p(t-T_p)e^{-i\omega (t-T_p)}
\end{equation}
 with the envelope function $f_p(t)$, 
the conductivity defined by
Eq. (\ref{td_sigma}) becomes
\begin{eqnarray}
\tilde \sigma(T,\omega)
&=& \frac{\int ds f(s) \tilde\sigma^I(T_p+s,\omega) e^{i(\omega-\omega_0)s}}
{\int ds f(s) e^{i(\omega-\omega_0)s}}
\nonumber\\
&=&
\sum_n \frac{\tilde f_p(\omega+n\Omega-\omega_0)}{\tilde f_p(\omega-\omega_0)}
e^{in\Omega T_p} \sigma^{(n)}(\omega+n\Omega),~~~~~
\label{td_sigma_general}
\end{eqnarray}
where $\tilde f_p(\omega)$ is the Fourier transform of $f_p(t)$.
In the case of linearly polarized light, $\tilde f_p(\omega)$ is the important parameter \cite{otobe16}.

\subsection{Parabolic two-band model}
We will introduce a two-band model, considering
only two orbitals in the sum the occupied valence ($v$) and unoccupied conduction
($c$) bands. The excitation energy from the valence band
to the conduction band is assumed to have a parabolic form,
\begin{equation}
\epsilon_{c \vec k} - \epsilon_{v \vec k}
\simeq \frac{k^2}{2\mu} + \epsilon_g,
\end{equation}
where $\epsilon_g$ is the band gap energy and $\mu$ is the reduced mass
of electron-hole pairs.

The time-dependent conductivity for parabolic two-band system can be written as
\begin{eqnarray}
\label{eq:sigma_c}
\sigma(T,\omega)&=&\int_{-\infty}^{\infty}ds e^{i\omega s}\sigma(T,T-s) \nonumber\\
&=&\frac{ie^2}{m_e\omega}n_e+\frac{e^2|p_{cv}|^2}{m_e^2\omega V}\int _0^{\infty} ds  e^{i\omega s}\nonumber\\
&\times&\sum_{\vec{k}}\Big[ e^{-i\int_0^s dy \left\{\frac{1}{2\mu} \left( \vec{k}+\frac{e}{c}\vec{A}(T-y)\right)^2+\varepsilon_g\right\}}\nonumber\\
&-& e^{i\int_0^s dy \left\{\frac{1}{2\mu} \left( \vec{k}+\frac{e}{c}\vec{A}(T-y)\right)^2+\varepsilon_g\right\}}\Big].
\end{eqnarray}
The integral in the exponential is calculated as 
\begin{eqnarray}
\label{eq:Phase}
&&\int_0^s dy \left\{\frac{1}{2\mu} \left( \vec{k}+\frac{e}{c}\vec{A}(T-y)\right)^2+\varepsilon_g\right\}\nonumber\\
&\equiv&\left(\varepsilon_g+\varepsilon_k+U_c\right)s \nonumber\\
&-& \theta_1\left\{ \sin(\Omega(T-s)-\phi)-\sin(\Omega T-\phi)\right\},
\end{eqnarray}
where $\theta$ ($\phi$) is the angle between the propagation direction ($x$-axis) and $\vec{k}$ .
Here we introduce
\begin{equation}
U_c=\frac{e^2A_0^2}{2\mu c^2},
\end{equation}
and
\begin{equation}
\theta_1=\frac{ekA_0\sin\theta}{\mu c\Omega},
\end{equation}
where $U_c$ is the kinetic energy of the electron-hole pair which corresponds to the ponderomotive energy
 $U_p=\frac{e^2A_0^2}{4\mu c^2}$ under a linearly polarized light. 

The Fourier transformation of Eq. (\ref{eq:Phase}) is written as
\begin{eqnarray}
&\int _0^{\infty}& ds  e^{i\omega s}\exp\left[-i \left(\varepsilon_g+\varepsilon_k+U_c\right)s \right. 
\nonumber\\
&-&\left.
\theta_1\left\{ \sin(\Omega(T-s)-\phi)-\sin(\Omega T-\phi)\right\}\right] \nonumber\\
&=&\sum_{l,l'}J_{l}(\theta_1)J_{l'}(\theta_1)\int_0^{\infty} ds  e^{i\omega s}\exp\left[ -i\left(\varepsilon_g+\varepsilon_k+U_c\right)s\right] 
\nonumber\\
&\times&
\exp[-il(\Omega(T-s)-\phi)+il'(\Omega T-\phi)]\nonumber\\
&=&i\sum_{l,m}J_{l}(\theta_1)J_{l+m}(\theta_1)\frac{\exp[im(\Omega T-\phi)]}{\omega-(\varepsilon_g+\varepsilon_k+U_c+ l\Omega)},
\label{Phase}
\end{eqnarray}
where $J_l$ is the $l$'th order Bessel function. In the last step in Eq.~(\ref{Phase}), we change $l'$ to $l+m$ to simplify the equation.

By substituting Eq. (\ref{Phase}) into Eq. (\ref{eq:sigma_c}), the $\sigma(T,\omega)$ can be expressed by the simple form,
\begin{eqnarray}
&&\sigma(T,\omega)=\frac{ie^2}{m_e\omega}n_e\nonumber\\
&+&\frac{ie^2|p_{cv}|^2}{m_e^2\omega V}\sum_{\vec{k},l,m}J_{l}(\theta_1)J_{l+m}(\theta_1)\nonumber\\
&\times&\left[ \frac{e^{im(\Omega T-\phi)} }{\omega-(\varepsilon_g+\varepsilon_k+U_c+ l\Omega)}\right. \nonumber\\ 
&-&\left. \frac{e^{-im(\Omega T-\phi)}}{\omega+(\varepsilon_g+\varepsilon_k+U_c+ l\Omega)}\right].
\end{eqnarray}
From the Fourier transformation of $\sigma(T,\omega)$, the component in Eq. (\ref{td_sigma}) is 
\begin{eqnarray}
&&\tilde{\sigma}^{(n)}(\omega)=\frac{1}{T_{\Omega}}\int_0^{T_{\Omega}} dt e^{-in\Omega t} \sigma(t,\omega)\nonumber\\
&=&\frac{ie^2}{m_e\omega}n_e\delta_{n,0}+\frac{ie^2|p_{cv}|^2}{m_e^2\omega V}\sum_{\vec{k},l}\nonumber\\
&\times&\left[ \frac{J_{l}(\theta_1)J_{l+n}(\theta_1)e^{-in\phi} }{\omega-(\varepsilon_g+\varepsilon_k+U_c+ l\Omega)} \right. \nonumber\\
&-&\left. \frac{J_{l}(\theta_1)J_{l-n}(\theta_1)e^{in\phi}}{\omega+(\varepsilon_g+\varepsilon_k+U_c+ l\Omega)}\right].
\end{eqnarray}
Now, we have the time-dependent conductivity, 
\begin{eqnarray}
\label{SUMK}
&&\sigma(T,\omega)=\sum_n e^{in\omega T}\tilde{\sigma}^{(n)}(\omega+n\Omega)\nonumber\\
&=&\frac{ie^2}{m_e\omega}n_e+\sum_{\vec{k},l,n}\frac{ie^2|p_{cv}|^2}{m_e^2(\omega+n\Omega) V}e^{in\omega T}\nonumber\\
&\times&\left[ \frac{J_{l}(\theta_1)J_{l+n}(\theta_1)e^{-in\phi} }{\omega+n\Omega-(\varepsilon_g+\varepsilon_k+U_c+ l\Omega)} \right.\nonumber\\
&-&\left.
\frac{J_{l}(\theta_1)J_{l-n}(\theta_1)e^{in\phi}}{\omega+n\Omega+(\varepsilon_g+\varepsilon_k+U_c+ l\Omega)}\right].
\end{eqnarray}
We can change the sum over $\vec{k}$ into integral by
\begin{equation}
\frac{1}{V}\rightarrow\frac{\mu^{3/2}}{2\sqrt{2}\pi^3} \int_0^{\infty} \sqrt{\varepsilon_k}d\varepsilon_k\int _{-1}^{1} d\cos\theta \int_0^{2\pi} d\phi
\end{equation}
Then the Eq. (\ref{SUMK}) has the form, 
\begin{eqnarray}
\label{eq:Cir_DFKE}
&&\sigma(T,\omega)=\frac{ie^2}{m_e\omega}n_e+\sum_{l,n}\frac{ie^2|p_{cv}|^2\mu^{3/2}}{\sqrt{2}\pi^2 m_e^2\omega}\nonumber\\
&\times&e^{in\omega T}\int_0^{\infty} \sqrt{\varepsilon_k}d\varepsilon_k\int _{-1}^{1} d\cos\theta J^2_{l}(\theta_1) \nonumber\\
&\times&\left[ \frac{\delta_{n,0}}{\omega-(\varepsilon_g+\varepsilon_k+U_c+ l\Omega)} \right. \nonumber\\
&-&\left.
\frac{\delta_{n,0}}{\omega+(\varepsilon_g+\varepsilon_k+U_c+ l\Omega)}\right].
\end{eqnarray}
The delta function, $\delta_{n,0}$, which is derived from the integration  about the angle $\phi$, omit the time-dependent oscillation.
Thus, the time-dependent conductivity becomes time-independent function. 
As the final result, we obtain the following expression for the real-part of the conductivity, 
\begin{eqnarray}
\label{eq:Cir_DFKE_Re}
{\rm Re} \sigma(\omega)&=&\sum_{l}\frac{e^2|p_{cv}|^2\mu^{3/2}}{\sqrt{2}\pi^2 m_e^2\omega}\nonumber\\
&\times&(\xi_l(k^+)\sqrt{\epsilon^+_{k,l}}-\xi_l(k^-)\sqrt{\epsilon^-_{k,l}}), 
\end{eqnarray}
where 
\begin{equation}
\xi_l(k)=\int _{-1}^{1} d\cos\theta J^2_{l}(\theta_1)
\end{equation}
and $k$( $k^{\pm}$) is related to $\epsilon_k$ 
($\epsilon_k^{\pm}$) by $k=\sqrt{2\mu \epsilon_k}$.
$\epsilon_k^{\pm}$ are defined by
\begin{equation}
\epsilon^{\pm}_{k,l}=\pm\omega-(\epsilon_g+U_c+l\Omega).
\end{equation}
Under a linearly polarized field, the Fourier component of the probe field in Eq.~(\ref{td_sigma_general})
also controls the response.
However, under a circularly polarized light, this probe field dependence disappears.

The conductivity is connected to the dielectric function by the relationship
\begin{equation}
\varepsilon(\omega)=1+i\frac{4\pi}{\omega} \sigma(\omega).
\end{equation} 
The important feature of Eq.~(\ref{eq:Cir_DFKE}) is the absence of the time-dependence, 
in contrast to the linear polarized case \cite{otobe16}.
The time-dependent change of the optical properties under linear polarized light is due to the response of the  different dressed state.
In the case of circularly polarized light, this effect disappears due to the integration about $\phi$ that corresponds to the average of the time dependence.
In real system, since the reduced mass depends on the $\theta$ and $\phi$,
 the time-dependence by the angle dependence of $\mu$ should appears. 
 
\subsection{Comparison between circularly and linearly polarization}
We will now revisit the Tr-DFKE with linearly polarized light for the purpose of comparison.
The real-part of the time-dependent conductivity ${\rm Re} \tilde{\sigma}^l(T_p,\omega)$ is expressed as,
\begin{eqnarray}
\label{sigma_CS}
&&{\rm Re} \sigma^L(T_p,\omega)
= 
 \sum_{m=-\infty}^{\infty}\frac{e^2 \mu^{3/2}\vert (p_{\alpha})_{vc} \vert^2}{\sqrt{2} \pi^2 m_e^2 (\omega+2m\Omega)} 
\nonumber\\
&&\times\frac{\tilde f_p(\omega + 2m\Omega-\omega_0)}{\tilde f_p(\omega-\omega_0)}\big[
 C_m(\omega)  \cos 2m\Omega T_p \nonumber\\
 &&+ S_m(\omega) \sin 2m\Omega T_p\big].
\end{eqnarray}
Coefficients $C_m(\omega)$ are given by
\begin{eqnarray}
C_m(\omega)=
\sum_l \pi \left[ 
\sqrt{\epsilon_k^{L,+}} \xi_{l,2m}^L(k^+)
-
\sqrt{\epsilon_k^{L,-}} \xi_{l,-2m}^L(k^-) \right],~~
\label{Cm}
\end{eqnarray}
where $\epsilon_k^{L,\pm}$ are defined by
\begin{equation}
\epsilon_k^{L,\pm} =
\pm (\omega + 2m\Omega) - (\epsilon_g + U_p +l \Omega),
\end{equation}
and 
\begin{equation}
 \xi_{l,2m}^L(k)=\int_{-1}^1 d(\cos\theta^L) J_l(\alpha,\beta)J_{l-2m}(\alpha,\beta).
\end{equation}
Here, $J_l(\alpha,\beta)$ is the generalized Bessel function \cite{otobe16,Reiss03},  $\theta^L$ is the angle between polarization direction and $\vec{k}$, 
\begin{equation}
\alpha=\frac{e k A_0\cos\theta^L}{\mu c \Omega},
\end{equation}
and
\begin{equation}
\beta=\frac{e^2 A_0^2}{8\mu c^2 \Omega}.
\end{equation}
The coefficients $S_m(\omega)$ are given by
\begin{eqnarray}
&&S_m(\omega)
=
-\int_0^{\infty} \sqrt{\epsilon_k} d\epsilon_k \nonumber\\
&\times&\sum_l \Bigg[
\frac{\xi_{l,2m}^L(k)}{\omega+2m\Omega-(\epsilon_k+\epsilon_g+U_p+l\Omega)} \nonumber\\ 
&&+\frac{\xi_{l,-2m}^L(k)}{\omega-2m\Omega+(\epsilon_k+\epsilon_g+U_p+l\Omega)} \Bigg].
\label{Sm}
\end{eqnarray}

 We note that although the Tr-DFKE under the circularly polarized light depends on the two parameters, $\theta_1$ and $U_c$, 
 the Tr-DFKE under the linearly polarized light depends on three, $\alpha$, $\beta$, and $U_p$.
 In particular, $\beta$ is the ratio between the ponderomotive energy and the photon energy of the pump laser which
 corresponds to the  adiabatic parameter, $\gamma=U_p/\Omega$ \cite{Nordstorm98}.
 In the case of the circularly polarization, $\gamma$ is not included in Eq. (\ref{eq:Cir_DFKE_Re}).  
 On the other hand, the $eA_0/\mu c\Omega$ appears in Eq. (\ref{eq:Cir_DFKE_Re}) and Eq.(\ref{sigma_CS}) .
\section{Numerical results}

\begin{figure}
\includegraphics[width=90mm]{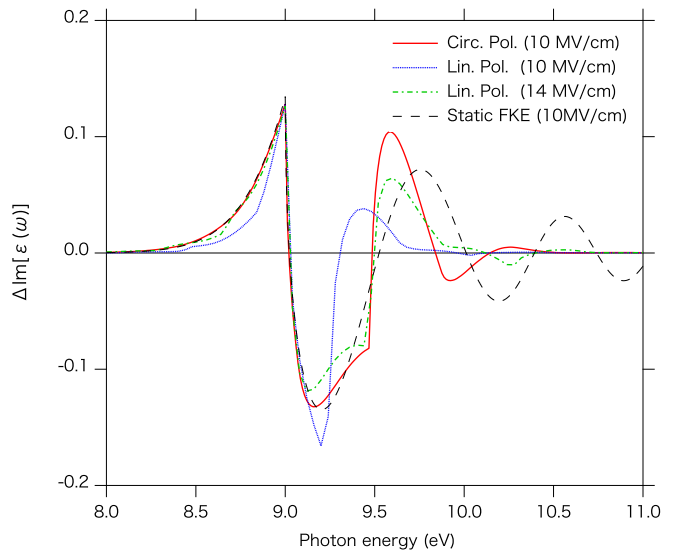}
\caption{\label{fig:Fig1} Change in ${\rm Im}[\varepsilon(\omega)]$ under the coherent light field with frequency of 0.4 eV and a field intensity 10 MV/cm.
(Red solid line) The $\Delta {\rm Im}[\varepsilon(\omega)]$ under  circularly polarized light. (Blue dotted line) Time averaged $\Delta {\rm Im}[\varepsilon(T, \omega)]$ 
under  linearly polarized light. (Green dash-dotted line) Time averaged $\Delta {\rm Im}[\varepsilon(T, \omega)]$ 
under  linearly polarized light with a peak field intensity of 14 MV/cm. (Black dashed line) The static FKE \cite{Tharmalingam63}.}
\end{figure}

In this section we aim to demonstrate the quantitative difference between circular and linear polarization 
by comparing the numerical results.
$\alpha$-quartz is a typical dielectrics used in non-linear laser-matter interaction studies, and we   
selected it as an example with which to illustrate the application of the foregoing formalism.
We assumed that the band gap, $\varepsilon_g$, was 9 eV,  the effective mass $\mu$ was 0.5$m_e$, and the transition moment $P_{cv}$ was one.

Figure (\ref{fig:Fig1}) show the change of the imaginary part of the dielectric function, $\Delta {\rm Im}[\varepsilon(\omega)]$.
The maximum field intensity was 10 MV/cm and the $\Omega$ was 0.4 eV.
The red solid line presents the circularly polarized light.
The exponentially tail below the band gap, which is one of the feature of the Franz-Keldysh effect, can be seen.
The oscillation above the band gap is owing to the absorption by the different $l-$th dressed states and the blue shift of the band gap by $U_c$.  
In this case, the  $U_c$ is approximately 0.5 eV which corresponds to the energy at which $\Delta {\rm Im}[\varepsilon(\omega)]$ becomes zero.

The time-averaged DFKE under linearly polarized light is presented by the blue dotted line.
The overall features are quite different to the circularly polarized case.
The most obvious difference is the amount of the blue shift, since the ponderomotive energy  ($U_p$)
under the linearly polarized light is half of the $U_c$.

At the same field intensity, the photon density for circularly polarized light is twice that for linearly polarized light. 
The time-averaged DFKE under linearly polarized light, whose photon density was equal to that of circularly polarized light,
 is presented by the green dash-dotted line. 
 In this case, not only the amount of blue shift, but also the behavior below the band gap is similar to the result with the circularly polarized light. 
 However, the oscillation above 9.5 eV is in contrast, with the difference possibly occurring due to the feature of each dressed state. 
 The intensities of dressed states are expressed by the generalized Bessel function for linearly polarized light. 
 On the other hand, for circularly polarized light, the dressed state is expressed by the Bessel function.


As a reference, the numerical result for the static FKE \cite{Tharmalingam63} is represented by the blak dashed line.
Since the applied field intensity is stable with circular polarization, the tunneling effect is expected to be similar to the static FKE. 
As expected, the behavior under the band gap agrees with the case of the circularly polarized light.
However, the oscillation above the band gap is different.
This result indicates that the FKE under the circularly polarized light is not  equivalent to the static FKE,
 even though the intensity of the electric field is static.
Thus, the interpretation by the dressed states is indispensable.  
\section{Summary}
We present the analytical DFKE formulation for circularly polarized light.
We found that the time-dependent change of the optical properties observed under the linearly polarized light disappears. 
Therefore, the linearly polarized light is suitable for the ultrafast control of the material response.
On the other hand, the response of the dressed state is still important for understanding the change in the optical properties caused by a circularly polarized light field.
We also find that the $\gamma=U_p/\Omega$ is not important parameter for the circularly polarization case, 
but the value $eA_0/\mu c\Omega$ is important for both of the circular and linear pump light.
\section*{Acknowledgement}
This work is supported by a JSPS KAKENHI (Grants No. 15H03674). 
Numerical calculations were performed on the supercomputer SGI ICE X at 
the Japan Atomic Energy Agency (JAEA).

\end{document}